# The AI-Augmented Research Process

# A Historian's Perspective

**Christian Henriot**
*Aix-Marseille University*

**Abstract[1]**

This paper presents a detailed case study of how artificial intelligence, especially large language models (LLMs), can be integrated into historical research workflows. The workflow is divided into nine steps, covering the full research cycle from question formulation to dissemination and reproducibility, and includes two framing phases that address setup and documentation. Each research step is mapped across three operational domains: (1) "LLM," referring to tasks delegated to language models; (2) "Mind," referring to the historian's conceptual and interpretive contributions; and (3) "Computational," referring to conventional programming-based methods (Python, R, Cytoscape, etc.). The study emphasizes that LLMs are not replacements for domain expertise but can support and expand historians' capacity to process, verify, and interpret large corpora of texts. At the same time, it highlights the necessity of rigorous quality control, cross-checking outputs, and maintaining scholarly standards. Drawing from an in-depth study of three Shanghai merchants, the paper also proposes a structured workflow tbased on a real case study hat articulates the historian's cognitive labor with both computational tools and generative AI. This paper makes both a methodological and epistemological contribution by showing how AI can be responsibly incorporated into historical research through transparent and reproducible workflows. It is intended as a practical guide and critical reflection for historians facing the increasingly complex landscape of AI-enhanced scholarship.

[1] I wish to thank Cécile Armand (CNRS) for her stimulating and incisive comments on an earlier draft of this paper. Her suggestions have been instrumental in shaping the revised version.

# Introduction

The impetus for this preliminary reflection on the use and impact of AI in historical research was the recent *AI in Science – Stakeholders Online Workshop* organized by the [European Commission's Directorate-General for Research and Innovation](#) (DG RTD) and the [Joint Research Centre](#) (JRC) on May 15, 2025. In the opening presentation, participants were shown a slide summarizing the "Scientific Process" (Fig. 1), intended to prompt discussion on where and how AI contributes at each stage. My immediate reaction was that this model—designed primarily for the natural and experimental sciences—did not capture the more diverse research processes characteristic of the social sciences and humanities, which operate in a different ecosystem, particularly in terms of publication and dissemination practices. Moreover, it failed to acknowledge the complex, iterative dynamics that occur between human cognition, computational methods, and large language models (LLMs).

**Figure 1. The Scientific Process**

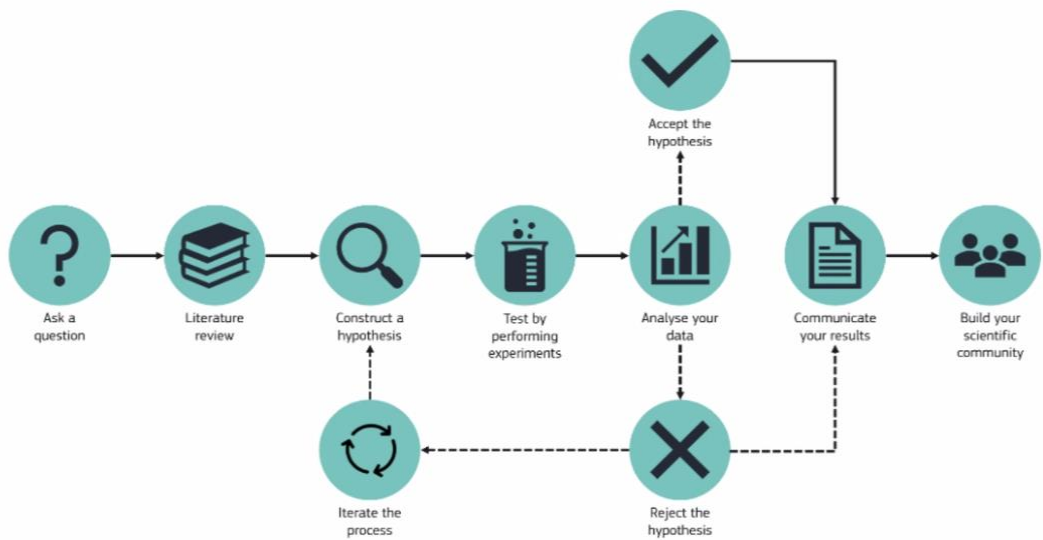

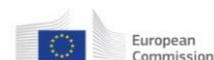

What historians—and other humanists—now confront is not entirely unexpected. They are, in fact, intellectually well-equipped to deal with these challenges, even if many (if not most) have chosen to disregard or sideline them. From the moment we had access to JSTOR, for instance, we should have asked: Who has access? What is included or excluded? How do we explore or read millions of pages? Our students were quick to draw their own conclusions: a journal is either available online, or it might as well not exist. When I began teaching my seminar "History in the Digital Age" in 2010, the focus was not on "the digital"—and certainly not on "digital history." It was about historical methods: expanding the historian's toolbox and broadening our critical horizon. It was also about the imperative to follow technological developments—especially those applicable to our sources—and to adopt any methodology that aligned with our research interests and materials.

Instead, we found ourselves overwhelmed by buzzwords—digital history, digital humanities, spatial humanities—that created unnecessary divides and tribes among humanists: between the "brave new world" and the guardians of humanistic tradition.[2]

Yet resisting the tide did not shield historians from the waves that followed. On one hand, the rapid digitization of sources, combined with their conversion into full-text searchable corpora—newspapers, periodicals, directories, even archival material—has created a research ecosystem that exceeds the capacity of human cognition to access, read, and analyze these materials using close-reading methods. Looking further ahead, the vastly larger digital archives being generated by governments, agencies, corporations, and individuals (through social media, websites, etc.) since the early 2000s pose an even greater challenge. Unless historians acquire the skills necessary to navigate this complex and uncharted terrain, this "world of abundance" will remain beyond their reach.[3] Historians of China may still feel secure working with Qing archives, Republican-era materials, or genealogies—but they must prepare the next generation to study Hu Jintao's or Xi Jinping's China. This is the reality that confronts historical scholarship *today*.

The ENP-China project was conceived and designed to confront this methodological challenge directly: how to retrieve historical information from vast digital corpora not merely through data mining, but through text analysis and machine learning, and how to process the resulting data into structured datasets at scale. The transition from scanned image-based texts to machine-readable corpora required historians to adopt computational methods in order to navigate and interpret these vast new textual resources. We embraced programming languages and even developed new tools—such as HistText, —specifically tailored for historical research. Building on

---

that experience, I published a blog post five years ago titled "[Rethinking historical research in the age of NLP](#)", which explored the transformative impact of digital technologies on the field, with particular attention to the integration of [Natural Language Processing](#) (NLP) techniques in historical research. In that piece, I identified three major challenges: the reliance on technically complex tools, the overwhelming scale of extracted data, and the difficulty of bridging quantitative analysis with qualitative historical interpretation.

I argued that NLP could significantly enhance historical inquiry by enabling large-scale extraction of entities and patterns from texts. However, it also introduced new methodological complexities. Historians needed to reconfigure their workflows and build supporting infrastructures to preserve the integrity of source-based analysis while leveraging computational power. I proposed a five-step protocol for processing historical corpora: (1) segmentation of raw documents, (2) indexing, (3) query and data extraction, (4) data exploration and cleaning, and (5) compilation and preservation. While NLP allows broader access to sources, it does not replace interpretation. Rather, it requires a deeper methodological integration between historians and computational tools to manage complexity and uphold scholarly rigor. The insights gained from this experience led us to rethink more radically our practices and, almost two years ago, we argued to go beyond digital humanities and take a decisive step toward computational methods.[4]

The rise of large language models (LLMs)—now often, though inaccurately, conflated with "AI"—has introduced a new paradigm, largely due to their promise of seamless and user-friendly integration into humanities research, bypassing the steep learning curve traditionally associated with programming languages. This more accessible pathway, however, carries renewed risks—not because LLMs are inherently flawed (though some raise ethical concerns), but because they are easily misused. One reason for such missteps lies in the common conflation of "AI" with the latest generation of LLM-based chatbots. Artificial Intelligence is not a new field; it encompasses numerous subfields, with Generative AI being merely the most recent development (Figure 2).[5] Yet we are once again falling into the familiar trap of embracing a new buzzword wholesale, conjuring new demons and fanciful imaginaries.

Unlike digital humanities, which largely remained within scholarly circles, AI is penetrating deeply into both society and historical research—from popular media representations to interventions by computer scientists.[6] At this critical juncture, I contend that the most urgent question is not whether AI will replace historians, or what AI *can do for* historians, but rather what *historians can do with* AI.

---

[4] Cécile Armand and Christian Henriot, "Beyond Digital Humanities Thinking Computationally: A Position Paper," 2023, https://shs.hal.science/halshs-04194570.

[5] All visualizations and tables are available on my [GitHub repository](#).

In this paper, I proceed in three steps. First, I discuss what Artificial Intelligence (AI) represents as a field of knowledge and how it has evolved historically. Second, I introduce an AI-augmented workflow for historical research, presented as an ideal type in nine steps, with a detailed description of what large language models (LLMs) can contribute at each stage. Third, I offer a concrete case study drawn from my own research to demonstrate more complex levels of interaction between mind and machine.

**Figure 2. The World of Artificial Intelligence**

Source: [Wikipedia Commons](...) (left); (right) [Datasciencedojo](...)

AI has a long and complex history, with roots that can be traced back to the 17th century in Descartes' logical frameworks and to the 19th century with the invention of the Jacquard Loom and Lovelace's algorithm (Figure 3). More direct developments occurred in the 20th century, including the Turing Machine (foundational to computation theory), early neural network models, and AI-based game programs. The pivotal moment, however, came in 1956, when twenty scientists convened at the

---

[Dartmouth Math Department   for a Summer Research Project](#)—an event that effectively marked the formal establishment of AI as a distinct field.[7]

---

[7] For this section and Figure 2, I have drawn extensively from Baptiste Blouin's presentations at EHESS (December 2024) and the Institute of Modern History (January 2025).

**Figure 3. The Evolution of Artificial Intelligence**

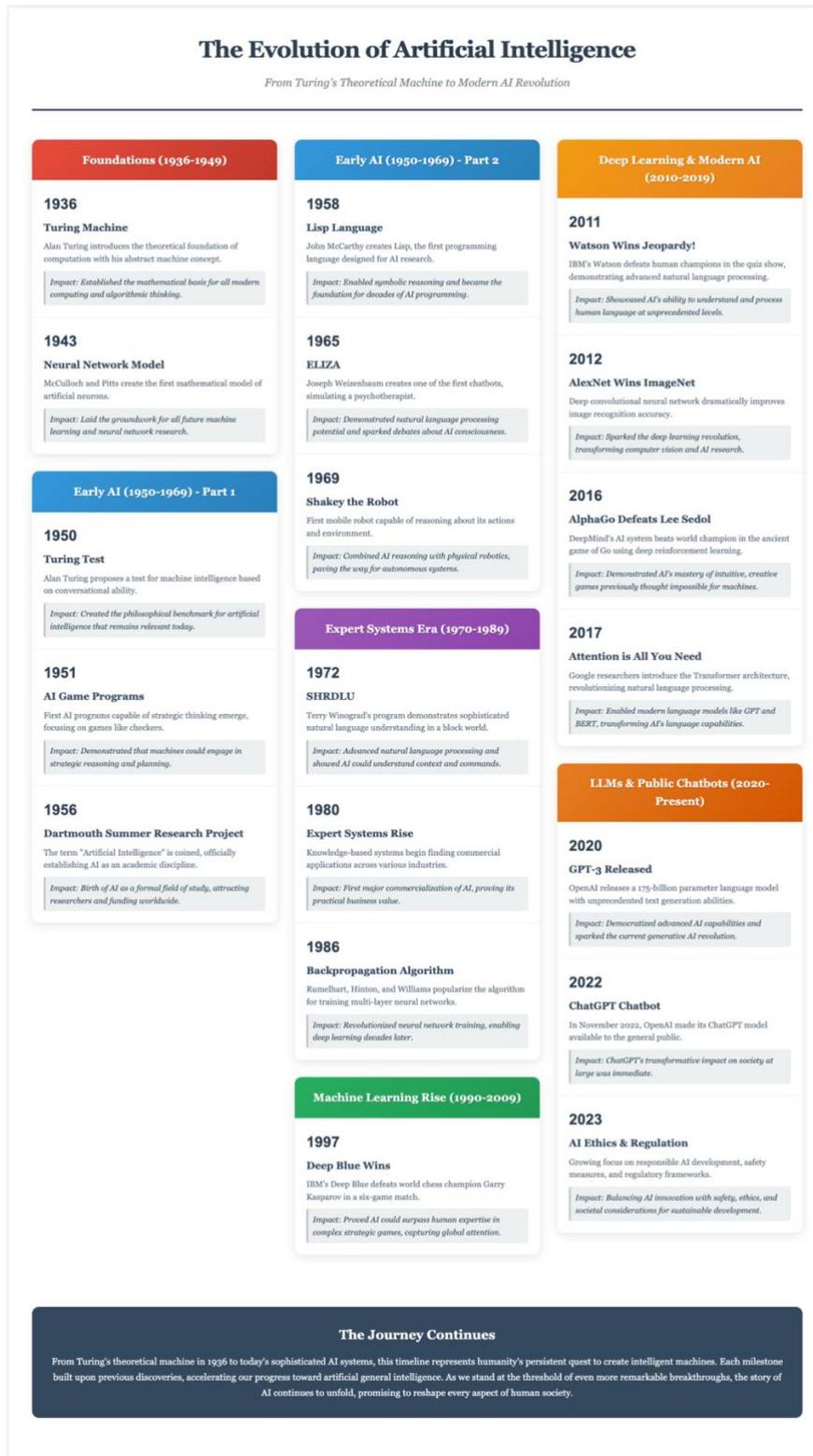

Note: Graph visualization produced with Claude Sonnet 4

[Click here to see the online interactive version](#).

Understanding this broader historical and conceptual landscape is essential, not only to avoid conflating distinct technologies but also to approach LLMs with the necessary clarity and precision. With this foundation in place, let me now turn to the practical conditions under which LLMs can be productively integrated into humanities research. A relevant and fruitful use depends on carefully and aptly curated prompts, which in turn implies understanding how to turn complex historical questions into tasks that the LLM can properly interpret. If LLMs can indeed facilitate access to complex operations without coding, there are at least three significant benefits to learning a programming language: (1) it helps one understand how computers process instructions and structure tasks, which in turn facilitates more precise and effective prompt engineering; (2) many common NLP tasks do not require the power of LLMs and can be performed as or more efficiently using traditional machine learning tools and libraries in Python or R; and (3) programming proficiency confers greater autonomy and control over text and data processing workflows. For these reasons, I continue to advocate strongly for acquiring basic coding skills to enable the productive and responsible use of LLMs in historical research.

In the following section, I propose a new nine-step protocol for AI-augmented historical research. It reflects my own experience working with LLMs since early 2023, during which time I have witnessed remarkable advances across platforms such as OpenAI, Anthropic, DeepSeek, Gemini, and Mistral, as well as the proliferation of specialized AI tools for scholarly research (e.g., Perplexity, AfforAI, ResearchRabbit, ConnectedPapers, Elicit). This protocol is preceded by a brief discussion of the ethical considerations involved in using AI in historical scholarship. A workflow chart (Figure 4) —The A.I.-augmented Historical Research Process— visualizes the nine steps and the forms of AI support applicable at each stage. In the final section, I develop a case study workflow based on my own research to demonstrate that in an AI-augmented research process, historians still take center stage in an iterative process that combine LLMs, human cognition, and other computational methods. This paper intends to serve a dual purpose: to offer a critical reflection on the use of AI in historical research while also functioning as a practical, step-by-step guide—complete with concrete examples—on how to harness AI tools effectively throughout the research process.

The integration of artificial intelligence into historical research offers exciting possibilities, but it also demands a careful and principled approach. In line with the European Commission's Guidelines for Trustworthy AI, historians must adopt these technologies with a strong commitment to responsibility, transparency, and scholarly integrity. AI tools are not neutral; they carry built-in assumptions and limitations that can introduce distortions if not critically assessed. Human oversight remains essential at every stage.

Historians should manually verify any AI-generated output, whether it be translations, summaries, or metadata extraction. This is particularly important when dealing with multilingual sources, where cultural nuance and historiographical context are often lost or oversimplified. The same care applies to named entities, as variations in transliteration, institutional titles, or date formats can lead to misleading interpretations. Summaries and paraphrases, meanwhile, may omit critical qualifiers, shift tone, or erase the specificity of an archival record. Whatever the format, historians are ultimately responsible for the interpretive validity of their arguments—even when those arguments draw on AI-processed material.

Transparency in the use of AI tools is increasingly recognized as a scholarly obligation. Historians should clearly acknowledge the role of these tools in their work, whether through footnotes that cite the model and version (for example, "Draft abstract generated with assistance from GPT-4, OpenAI, reviewed and revised by the author"), prefaces or appendices noting the use of AI in translation or data extraction, or supplementary documentation that outlines how datasets or corpora were processed. Such transparency not only supports reproducibility but also affirms the place of AI tools as part of the historian's research infrastructure—not as hidden collaborators.

At the same time, the known limitations of large language models must not be overlooked. These models can generate hallucinated facts, invented references, and inaccurate citations, particularly when prompted to summarize or simulate academic literature. Historians should therefore approach AI-generated references with critical awareness, avoid using AI to reproduce secondary literature unless the original texts are available for verification, and make clear when examples are hypothetical—especially in teaching or public-facing work.

Upholding these principles will help ensure that AI serves as a meaningful aid to historical research, rather than a shortcut or source of error. It is only through critical engagement and clear attribution that AI can be responsibly integrated into the discipline.

**Understanding AI's Role in the Historical Workflow**

Throughout this workflow, Artificial Intelligence (AI) tools are positioned as *assistants* rather than interpreters (Figure 4). To clarify how AI contributes across the stages of research, I adopt a typology of five core functions:

- **Discover**: Identify patterns, gaps, or underexplored themes in corpora, archives, or scholarship.
- **Analyze**: Extract structured information, compare texts, cluster documents, or detect semantic trends.
- **Support Writing**: Generate, revise, or clarify prose; structure arguments; test coherence.
- **Visualize**: Transform data, timelines, or networks into interpretable maps, charts, or diagrams.
- **Translate & Contextualize**: Render foreign-language materials into usable summaries, and assist with interpreting culturally specific terms or frameworks.

## Color Code of AI Function Typology

| AI Function | Color | Justification / Association |
|---|---|---|
| **Discover** | 🔵 **Blue** | Exploration, knowledge discovery, search |
| **Analyze** | 🟠 **Orange** | Processing, dissection, pattern recognition |
| **Support Writing** | 🟢 **Green** | Composition, construction, intellectual growth |
| **Visualize** | 🟣 **Purple** | Diagrams, abstraction, transformation of structure |
| **Translate & Contextualize** | 🔴 **Red** | Bridging meanings, decoding cultural signals |

These categories serve as a shorthand for the AI affordances described at each step. They are not rigid: many tools perform multiple functions (e.g., GPT-4 can support both translation and argument development). However, this typology helps foreground the *cognitive and methodological diversity* of AI use in historical practice, without flattening the historian's interpretive labor.

**Figure 4. The AI-Augmented Historical Research workflow**

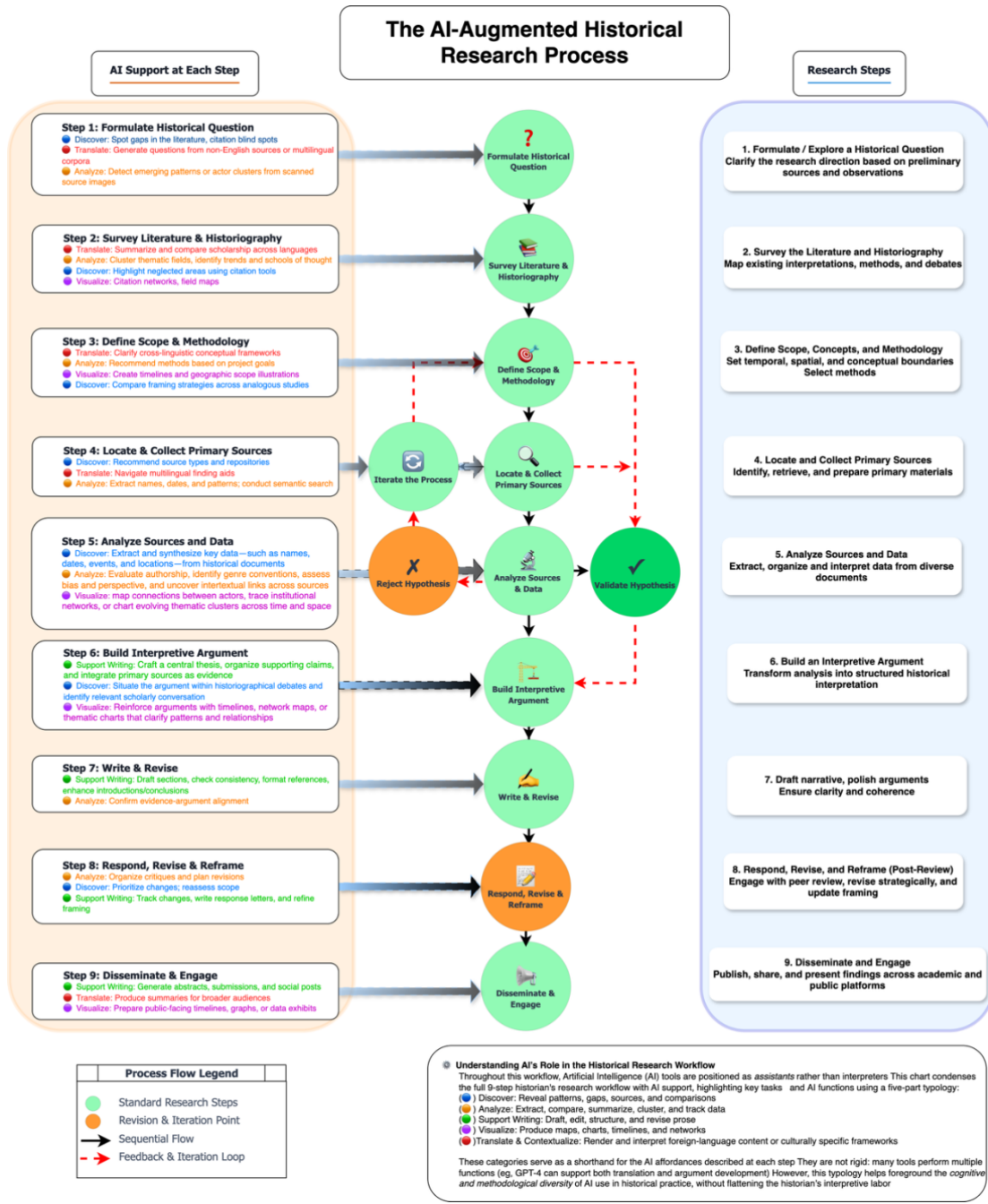

Note: Graph visualization produced with draw.io.

Click on this link to see the image online.

# Step 1: Formulate/Explore a Historical Question

 **Description**

Identify a meaningful, researchable question grounded in historical context. This may arise from a historiographical puzzle, newly available sources, a methodological innovation, comparative curiosity, or preliminary work on historical sources. This step is exploratory and iterative. Questions are tentative, revised in light of reading, archival leads, or even methodological shifts.

*Example*: "What role did merchant networks play in philanthropic and public health initiatives in treaty-port Shanghai during the 1920s?"

 **AI's Role**

🔵 *Discover* - **1. Identify Gaps or Underexplored Themes**

- ▪ **How**: Use NLP techniques like *topic modeling*, *named entity recognition*, or *citation network analysis* on large corpora.
- ▪ **Where**: Digitized journal databases (e.g., *JSTOR*, *Project MUSE*), specialized bibliographies (*Historical Abstracts*, *Bibliography of Asian Studies*), or full-text repositories (*HathiTrust*, *Google Books*, or *CAJ* for Chinese).
- ▪ **What this reveals**: 1. Absences or low-density areas in topical coverage; 2. Repetition of certain keywords without analytical development; 3. Historiographical saturation vs. blind spots.

*Example 1* : A topic model across 500 articles on "Shanghai merchants" from the press might show that philanthropy is often mentioned in passing but rarely studied as a structured phenomenon—signaling a potential research niche.

*Example 2*: While not full-text, *Historical Abstracts* can be mined for metadata trends (topics, regions, periods) over time through the metadata and abstracts, showing which areas have seen declining or surging scholarly attention.

🔴 *Translate & Contextualize* – **2. Generate Question Prototypes from Preliminary Source Snippets**

- • **What counts as "preliminary data"**: 1. A set of archival finding aids; 2. An early batch of digitized newspaper articles (e.g., *Shenbao*); 3. A memoir, institutional report, or dataset (e.g., biographical lists, gazetteers); 4. An export of search results from digital libraries.

- **AI techniques**: 1. Entity clustering (e.g., recurring actors in institutional reports); 2. Temporal trends (e.g., emergence or decline of topics in press over decades); 3. Co-occurrence mapping (e.g., links between "merchants" and "medicine").
- **What this reveals**: 1. Absences or low-density areas in topical coverage; 2. Repetition of certain keywords without analytical development; 3. Historiographical saturation vs. blind spots.

*Example*: Feeding 100 *Shenbao* articles into an LLM or NER pipeline could reveal repeated co-location of "Zhu Baosan," "hospitals," and "public subscriptions," prompting questions about the structure of medical philanthropy.

## 🔵 *Discover* – 3. Suggest Analogous Research Questions from Related Fields

**Method**: Train or prompt an LLM with a few example historical questions and ask it to suggest analogous ones based on thematic similarity, structural pattern, or comparative framing. This is especially useful for generating cross-regional comparisons, transperiodic inquiries, or counterfactual thinking.

**Example**: **Input**: Based on the question "How did Protestant missions reshape educational models in late Qing China?", generate three comparable research questions that vary either the religion, the social domain, or the regional context **AI Output**: "How did Buddhist charitable institutions reshape healthcare delivery in interwar Japan?"

**Explanation**: The AI recognizes the structure of the question—how a religious institution reshaped a domain of public life within a specific historical context—and replicates it by switching the religion (Buddhism), the domain (healthcare), the time period (interwar), and the geographic focus (Japan). This facilitates comparative or analogical thinking across contexts.

## 🔵 *Discover* – 4. Explore Research Trends via Citation Graphs

- **Tools**: Semantic Scholar, Connected Papers, OpenAlex
- **Function**: AI can highlight **clusters of scholarship** and their connections or isolation in the scholarly landscape.

*Example*: A visualized citation network of works citing Wellington K.K. Chan's work on "Changes in the Merchant's Roles, Class Composition, and Status" may reveal lateral themes (e.g., education, ) but an absence of work linking it to **civil society**.

🟠 *Analyze – 5. Generate Research Agendas from Digitized Source Corpora*

- **Sources**: Local gazetteers, missionary archives, newspaper corpora, parliamentary records.
- **AI Use**: Apply unsupervised topic modeling or semantic clustering to detect unexpected themes, geographical outliers, or policy inflection points.

*Example*: Running LDA (a topic modeling method) on British consular dispatches from Shanghai might uncover recurring concerns with Chinese-run hospitals—something that does not appear in existing historiography.

**AI Affordances in Step 1**

| Function | Input Required | Outcome for the Historian |
|---|---|---|
| Corpus-level gap detection | Journal archives, bibliographic metadata | Identify underexplored themes or historiographical niches |
| Entity/topic co-occurrence | Preliminary document sets | Spot patterns worth formalizing into research questions |
| Analogical question generation | Seed questions, disciplinary framing | Expand the horizon of inquiry through comparative prompts |
| Citation network exploration | Semantically linked databases (e.g., OpenAlex) | Understand the landscape and its blind spots |
| Early source corpus mining | Gazetteers, press articles, archival summaries | Detect leads from primary materials before full analysis |

Step 1 involves sophisticated techniques such as topic modeling or named entity recognition. However, historians without programming skills can still meaningfully leverage AI and large language models (LLMs), especially when tools are integrated into user-friendly platforms or accessed via natural language interfaces. While processing an entire corpus of hundreds or thousands of documents directly within a prompt window is not realistic, using an API offers a more practical solution. That said, historians **must first understand** what topic modeling entails in order to use it effectively. The same prerequisite applies to other digital methods, such as network analysis, sequence analysis, and beyond.

With a set of preliminary questions in place, the historian next turns to the existing scholarly terrain. What has already been said, how, and by whom? Step 2 focuses on surveying the relevant literature and historiography—not only to avoid duplication, but to position the research within established debates, identify interpretive trends, and reveal unexamined angles that AI tools can help illuminate.

# Step 2: Survey the Literature and Historiography

##  Description

Conducting a comprehensive review of existing scholarship serves not only to understand prevailing interpretations, influential methodologies, and dominant debates related to the research question, but also to identify gaps in coverage and opportunities for intervention. The aim is threefold: 1. to identify blind spots, neglected perspectives, or underexplored corpora; 2. to trace conceptual genealogies that have shaped current understandings; and 3. to situate one's own research within broader scholarly conversations—whether through thematic lenses (such as philanthropy, public health, or state-society relations), methodological frameworks (such as social or transnational history), or historiographical traditions. The latter may vary significantly across languages, institutions, and national contexts. In historically multilingual fields, engaging with literature in multiple languages becomes crucial. Each linguistic sphere is often shaped by distinct academic traditions, political constraints, and archival infrastructures, all of which must be taken into account to build a truly transnational or comparative historiographical foundation.

##  AI's Role

 **Translate & Contextualize** – 1. Multilingual Summarization and Comparison

- **What it does**: Translates and synthesizes arguments from academic works in different languages.

- **How**: 1. Use LLMs (GPT-4, Claude, Gemini) to summarize abstracts, introductions, or full-text passages in Chinese, Japanese, French, German, etc. 2. Compare methodological framing or terminology between linguistic corpora.
- **Use case**:
  - Input: abstracts or selected excerpts from French, Chinese, or Japanese monographs.
  - Output: cross-language summaries or thematic synthesis.

*Example 1*: "Summarize the methodology and main argument of this Japanese article on philanthropic institutions. How does it compare to Anglo-American approaches?"

*Example 2*: Paste Chinese abstracts from the CNKI database and ask, "What is the central argument and how does it compare to Western historiography on the same topic?"

*Practical*: A French-trained historian unfamiliar with Chinese can get high-level summaries of PRC scholarship.

## 🔴 *Analyze* – 2. Thematic Clustering and School Identification

- **What it does**: 1. Identifies clusters of texts by theme, approach, or school of thought; 2. Cluster articles by topic or argument type; 3. Suggest which ones are more methodologically innovative vs. empirically rich.
- **How**: 1. Use dedicated tools to group papers by shared keywords, cited works, or questions; 2. Generate overviews of *how* a topic has been studied—social history vs. institutional history vs. discourse analysis, etc.
- **Tools**: Elicit.org, ResearchRabbit, ConnectedPapers, Semantic Scholar, and in-development tools like Scite Assistant.

*Example*: Feed 10 articles on "Chinese merchant philanthropy" into Elicit.org to see which focus on economic theory, social networks, or institutional histories.

*Use case*: "Group literature on Shanghai business networks into economic, cultural, and political subfields."

## 🔵 *Discover* – 3. Build a Citation Map to Detect Gaps in the Literature

- **AI Contribution**: 1. Highlight isolated or under-cited works in other languages; 2. Suggest bridging works that cite across linguistic domains.

*Example*: A Chinese article frequently cited in PRC literature but absent from English bibliographies may suggest a blind spot in Anglophone scholarship.

**Tools**: OpenAlex, Semantic Scholar, Lens.org

- **What it does**: Highlights topics, actors, or regions that are underrepresented or methodologically neglected.
- **How**: 1. Run citation analyses via : OpenAlex or ConnectedPapers to detect which relevant works are **not cited** in dominant literature; 2. Use LLMs to analyze bibliographies and **flag absences** (e.g., "No works on women's roles in merchant philanthropy are cited").

*Example*: After surveying 30 English articles on Chinese philanthropy, ask: "What subtopics are consistently missing or only mentioned in passing?"

### 🔵 *Visualize* – 4. Visualize Citation Networks and Scholarly Influence

- **What it does**: Reveals intellectual lineages, isolated authors, or bridges between linguistic domains.
- **How**: 1. Tools like ConnectedPapers, OpenAlex, or Lens.org build citation graphs from seed articles; 2. AI can identify central vs. peripheral figures in a debate.

*Example*: Input a foundational article on late Qing philanthropy and map its intellectual descendants—and which schools of thought cite it.

### 🔴 *Translate & Contextualize* – 5. Translate and Contextualize Key Historiographical Concepts

- **What it does**: Helps historians interpret **culturally embedded terminology** across languages.
- **How**: LLMs can translate with conceptual sensitivity: e.g., not just "civilisation" → "文明" but also explaining connotations in Durkheimian vs. Confucian contexts.

*Prompt*: "Explain the difference between the French use of 'civilisation' and its Chinese equivalent 'wenming' in Republican-era discourse."

### Challenges

- **Limited database interoperability**: CNKI, Cairn, JSTOR, and other repositories often do not share metadata or citations.
- **Conceptual untranslatability**: Terms like *gongyi* (公益), *civilisation* (in French moral-political context), or *kyōka* (教化, moral reform) carry field-specific meanings.

**AI Mitigations**

- Use LLMs to translate not just words, but **frames of reference**:
  - "Explain what 公益 meant in Republican-era Chinese discourse."
  - "Translate and contextualize 'civilisation' as used in Durkheim's moral sociology."

AI isn't replacing human interpretation—it is *amplifying multilingual accessibility*.

**AI Affordances in Step 2**

| Function | Input Required | Outcome for the Historian |
|---|---|---|
| Multilingual summarization | Abstracts, excerpts, or full texts | Cross-linguistic understanding of arguments and methods |
| Thematic clustering | PDFs, article links, bibliographies | Identification of scholarly subfields and schools of thought |
| Gap detection | Citation metadata, bibliographies | Awareness of neglected regions, actors, methods |
| Citation mapping | Seed texts, DOIs | Visualized networks of influence and omission |
| Historiographical translation | Terms, passages, conceptual prompts | Contextualized understanding of discipline-specific terms |

Once the contours of the historiography are understood, the historian must define the boundaries of their own intervention. Step 3 involves shaping the project's conceptual, temporal, and methodological framework—decisions that will guide source selection, interpretive lens, and analytical tools. Here, AI supports clarity and coherence in framing, helping connect scholarly aims to workable research designs.

# Step 3: Define Scope, Concepts, and Methodology

 **Description**

This step involves transforming an exploratory question into a **workable research design**. Historians articulate:

- **Temporal scope** (e.g., 1905–1949 or Meiji to Shōwa),
- **Geographical framing** (local, regional, transnational),
- **Thematic and conceptual lenses** (e.g., "civil society," "statecraft," "gendered labor"),
- and **methodological orientation** (e.g., prosopography, discourse analysis, etc.).

This is not a mechanical narrowing of the topic, but an **intellectual act of framing** that structures the entire inquiry.

 **AI's Role**

🔴 *Translate & Contextualize* – **1. Assist in the Refinement of Concepts and Categories**

- **What it does**: AI helps historians clarify, compare, and sharpen conceptual frameworks.
- **How**: 1. Prompt LLMs with conceptual pairs (e.g., "charity" vs. "philanthropy," "reform" vs. "revolution") to explore historical meanings; 2. Generate typologies or definitional debates from existing scholarship.

*Example Prompt*: "Compare how Chinese-language and English-language historiography define 'public welfare' in early 20th-century Shanghai."
*Outcome*: Surface historiographical slippage, reveal where categories need revision.

🟠 *Analyze* – **2. Suggest Methodological Approaches Based on Research Goals**

- **What it does**: Connects research aims to plausible methods.
- **How**: 1. Given a project description, LLMs can list suitable methodologies (e.g., "You could consider social network analysis or institutional microhistory") ; 2. Provide annotated comparisons: e.g., differences between using *quantitative biography* vs. *actor-network theory*.

*Prompt*: "I am studying local hospital records and merchant associations in Shanghai—what methods might allow me to understand their interactions over time?"

🟣 *Visualize* – **3. Visualize Temporal and Spatial Boundaries**

- **What it does**: Offers graphic scaffolding for historical framing.
- **How**: 1. Use LLM-integrated tools to create timelines (e.g., Preceden, TimelineJS) or map affiliations across regions; 2. Identify chronological clusters in preliminary data (e.g., event spikes, publication surges).

*Example*: Upload a list of events or archival dates to generate a preliminary timeline of policy changes or philanthropic activity.

🔴 *Analyze* – **4. Help Formulate Operational Definitions and Source Selection Criteria**

- **What it does:** Assists historians in articulating clear definitions and selection criteria for identifying relevant material in a corpus.
- **How:**
  1. LLMs can support the formulation of inclusion criteria by helping refine conceptual boundaries. For example: "What qualifies as a 'merchant'? Should the category include foreign firms or only Chinese actors?"
  2. They can also help historians clarify distinctions between emic (actor-defined) and etic (analyst-defined) categories.

*Prompt*: "Help me define a workable set of criteria for identifying 'philanthropic institutions' in early 20th-century Shanghai press reports."

🔵 *Discover* – **5. Compare Framing Strategies Across Analogous Studies**

- **What it does**: Shows how similar projects elsewhere defined their parameters.
- **How**: Ask for examples from other historiographies: "How did French historians of the Third Republic define the social space of voluntary associations?"

- 

**AI Affordances in Step 3**

| Function | Input Required | Outcome for the Historian |
|---|---|---|
| Conceptual clarification | Key terms or binaries | Sharpened definitions and historiographical positioning |
| Methodological recommendation | Project description | Tailored methods aligned with research aims |
| Temporal/spatial visualization | Event lists, place names, periods | Framed scope for timelines or maps |
| Operational definitions | Research categories, draft questions | Clearer criteria for inclusion/exclusion of evidence |
| Framing analogies | Topic, region, comparative interest | Insights from related scholarly designs |

With a structured plan in hand, the task becomes operational: locating the sources that can bear interpretive weight. Step 4 marks the shift from project design to evidence gathering, including multilingual retrieval, transcription, and classification. AI proves especially valuable here in managing large digital archives and making diverse source types accessible for analysis.

# Step 4: Locate and Collect Primary Sources

 **Description**

This phase involves identifying, locating, and retrieving the primary sources that will form the evidentiary foundation of the research. These sources may include: 1. archival records from government, institutional, or private collections; 2. periodicals and newspapers; 3. pamphlets, gazetteers, and yearbooks; 4. photographs, maps, and oral histories; and 5. administrative reports, legal documents, and ephemeral materials. Historians undertaking this work must navigate several layers of complexity, including: 1. fragmented access across national and institutional boundaries; 2. variable formats such as print editions, digital scans, microfilms, or born-digital records; and 3. multilingual metadata or unindexed corpora that complicate discovery. The process requires both domain expertise and methodological flexibility to construct a robust and representative source base.

This stage is not purely logistical—it involves **strategic source thinking**: identifying which documents can speak to the research question, how silences operate, and how different genres might complement or contradict each other.

 **AI's Role**

🔵 *Discover* – **1. Identify Potential Source Types and Locations**

- **What it does**: Suggests relevant types of primary sources and where they might be found.
- **How**: 1. Given a research topic and period, LLMs can recommend source categories (e.g., tax rolls, guild minutes, orphanage reports) and known collections (e.g., Shanghai Municipal Archives, *North China Herald* corpus, etc.); 2. AI can also surface overlooked repositories or digital collections.

*Prompt*: "What types of primary sources might document philanthropic networks in 1930s Shanghai, and where might they be housed?"

🔴 *Translate & Contextualize* – **2. Assist in Navigating Archival Finding Aids and Catalogs**

- **What it does**: Helps interpret and summarize archival guides, especially in unfamiliar languages or formats.
- **How**: 1. OCR and translation tools can extract and render catalog entries from scanned PDFs; 2. AI can summarize or group entries thematically.

*Use case*: Translate and group all "慈善" (charity) entries from a Chinese municipal archive guide.

🟠 *Analyze* – **3. Extract and Clean Data from Digitized Sources**

- **What it does**: Prepares non-searchable or unstructured documents for analysis.
- **How**: This stage typically involves three main operations. First, scanned texts are processed using AI-powered optical character recognition tools such as Google Vision, Transkribus, or large language models with image input capabilities. Second, named entities, dates, and locations are extracted using pre-trained named entity recognition (NER) models, enabling structured exploration of unstructured texts. Third, long documents are segmented into logical units—for example, minutes, edicts, or reports—so that subsequent analysis can proceed with greater granularity and contextual precision.

*Example*: Process a 1935 *Shenbao* article to extract names of donors and their affiliations.

🟠 *Analyze* **– 4. Search Semantically Within Large Corpora**

- **What it does**: Goes beyond keyword search by retrieving texts by meaning, not exact phrasing.
- **How**: 1. LLM-based querying over corpora like the *North China Herald*, *Shenbao*, or *Dagongbao*—when interfaces permit (or via vector search models); 2. Ask natural-language questions: "Find reports about merchant-led relief efforts after the 1931 floods."

*Outcome*: Reduces missed documents caused by inconsistent terminology.

🔴 *Translate & Contextualize* **– 5. Translate and Summarize Foreign-Language Sources**

- **What it does**: Makes initial scanning and comprehension possible for non-native readers.
- **How**: 1. Use LLMs to translate entire documents or summarize their contents with attention to key people, places, and events; 2. Preserve ambiguity or signal uncertainty when present.

*Prompt*: "Summarize this 1947 Japanese municipal report and list all references to foreign relief agencies."

**AI Affordances in Step 4**

| Function | Input Required | Outcome for the Historian |
|---|---|---|
| Source type/location suggestion | Research topic, region, period | Targeted suggestions of document types and repositories |
| Finding aid navigation | Catalog text, PDFs, screenshots | Thematic organization, multilingual parsing |
| OCR + structured extraction | Scanned books, reports, images | Usable text, segmented content, named entities |
| Semantic document retrieval | Research question or theme | Context-relevant documents across corpora |
| Multilingual translation/summarization | PDFs, images, long texts | Comprehension of otherwise inaccessible materials |

At this point, a corpus of primary materials has been assembled. The next step is to read, contextualize, and critically interpret these sources. Step 5 is where historical judgment is most active: detecting bias, analyzing genre, mapping relationships, and extracting patterns. AI can assist in identifying consistencies or anomalies—but meaning and significance remain the historian's domain.

# Step 5: Source and Data Analysis

 **Description**

This step transforms raw source material into **interpretable historical evidence**. The historian must engage with documents on multiple levels:

- **Authorship and perspective**: Who created it, under what conditions, for whom?
- **Genre and form**: Report, speech, petition, editorial, etc.
- **Bias and silence**: What is included, excluded, emphasized, or concealed?
- **Embedded structure**: Who appears, when, in relation to what?

- **Intertextuality**: How does this relate to other documents?
- **Quantifiable content**: Extracting names, dates, places, affiliations, figures.

AI can assist in structuring, visualizing, and patterning the complexity of primary materials—but interpretive meaning must remain the historian's responsibility. Depending on the project, this may include **close reading**, **quantitative extraction**, **relational mapping**, or **discourse analysis**. AI can **support these tasks structurally**, but not interpretively.

**AI's Role:**

## 🔴 *Analyze* – 1. Contextualize Authorship and Provenance

**Goal**: Who created the source? For whom? Under what institutional or political conditions?

- **AI's Role**: 1. Extract and compile author metadata across multiple documents to establish patterns of authorship and contribution. 2. Apply named entity recognition (NER) to classify and group authors by institutional affiliation, geographic origin, or functional role. 3. Generate or retrieve biographical context using large language models or linked open data sources such as the China Biographical Database (CBDB) or Wikidata.

*Prompt*: "What do we know about the individuals named as authors in these reports from 1930s Guangzhou?" *Outcome*: Preliminary actor profiling.

## 🔴 *Analyze* – 2. Classify Genre, Form, and Intended Audience

**Goal**: Is this a report, petition, editorial, legal brief, minutes, speech? How does its form shape its content?

- **AI's Role**: 1. Use LLMs to classify genre based on document features; 2. Compare with genre typologies from known corpora (e.g., "What distinguishes these edicts from self-published pamphlets?").

*Use case*: Automatically flag documents as "statistical summary" vs. "narrative report" for batch analysis.

## 🔴 *Analyze* – 3. Assess Bias, Silences, and Representational Strategies

**Goal**: What is emphasized, omitted, repeated, or framed in a specific way?

- **AI's Role**: 1. Compare similar sources to detect lexical patterns, sentiment shifts, or recurring tropes; 2. Visualize term frequency or semantic proximity (e.g., "How often does 'merchant' appear alongside 'hygiene'?") 3. Use LLMs to simulate alternate framings ("Rewrite this passage from the perspective of a municipal doctor").

*Caution*: AI can detect patterns but not infer **why** they occur—that remains the historian's task.

## 🔴 *Analyze* – 4. Extract Structured Data Embedded in Text

**Goal**: Pull names, dates, places, numbers, categories from narrative text.

- **AI's Role**: 1. Named entity recognition for people, places, institutions; 2. Regex or LLM-based extraction of structured content (e.g., budgets, lists of donors, attendance logs); 3. Build event or actor databases from repeated entries.

*Example*: Extract names and affiliations of contributors to public health campaigns in a series of *Shenbao* articles.

## 🔵 *Visualize* – 5. Map Relational Patterns Across Documents

**Goal**: Identify social, institutional, or conceptual networks.

- **AI's Role**: 1. Generate co-occurrence networks to identify which individuals or entities appear together in texts and how frequently. 2. Apply unsupervised clustering algorithms to group documents by shared themes or associated actor sets. 3. Integrate extracted relational data into network visualization platforms such as Cytoscape or Palladio to explore structural patterns and centrality.

*Outcome*: Visualize the institutional network behind a series of philanthropic initiatives.

### 🔴 *Analyze – 6. Trace Intertextuality and Citational Patterns*

**Goal**: Understand how texts refer to, borrow from, or respond to each other.

- **AI's Role**: 1. Detect reused phrases, quotations, or structural mimicry across multiple texts. 2. Use embedding-based models to identify semantic echoes or paraphrased content between documents. 3. Summarize inferred intertextual relationships, such as identifying clusters of texts that reference or respond to a shared source (e.g., a 1931 ordinance).

### 🔴 *Analyze – 7. Support Discourse and Content Analysis*

**Goal**: Track how language encodes ideologies, institutions, or change over time.

- **AI's Role**: 1. Run topic modeling or dynamic word embedding models to identify evolving frames (e.g., how "hygiene" is used differently from 1910s to 1930s); 2. Segment discourses by speaker, topic, or publication source.

*Example*: Show how references to "modernity" shift from medical to architectural contexts over time.

##  AI Affordances in Step 5

| Task | AI Method / Tool | Output for the Historian |
|---|---|---|
| Author profiling and metadata | NER, biographical LLM queries | Biographical or institutional clusters |
| Genre and audience classification | LLM classification, prompt analysis | Source taxonomy by function or form |
| Bias and silence detection | Lexical frequency, comparison prompts | Highlighted emphases, omissions, discursive patterns |
| Data extraction from text | NER, regex, few-shot LLM prompts | Structured datasets (names, dates, numbers) |
| Relational mapping | Co-occurrence + network | Social or institutional graphs |

| | analysis | |
|---|---|---|
| Intertextuality tracing | Semantic similarity + citation mining | Maps of influence or repetition |
| Discourse evolution | Topic modeling, word embeddings | Shifting frames, terminologies, rhetorical anchors |

Once the interpretive skeleton is in place, it is time to bring the project to discursive life. Step 7 is the writing phase—where the historian drafts, refines, and polishes prose in a recursive and rhetorically sensitive process. AI can function here as an editorial assistant, supporting clarity, consistency, and narrative cohesion.

# Step 6: Build an Interpretive Argument

 **Description**

In historical scholarship, **building an interpretive argument** is where the historian's voice, framing, and originality become visible. It goes beyond simply "stating findings" to creating a **conceptually coherent, evidence-grounded, historiographically positioned** claim about the past. This step involves synthesizing evidence, historiography, and conceptual framing into a persuasive, structured historical argument. It is the moment where: 1. The problématique finds analytical expression. 2. Evidence is marshaled not just to illustrate, but to support and refine claims. 3. The interpretive stance of the historian becomes clear.

Rather than producing a simple linear narrative, this stage asks: 4. What is the central claim I am making about change, structure, agency, or meaning? 5. How do my sources, and their arrangement, support this? 6. How does my argument enter into conversation with or challenge existing scholarship? 7. Assume you are a historian who disagrees with this position—what might you say?

**AI's Role:**

🟢 *Support Writing* **– 1. Formulate the Core Interpretive Claim**

**Goal**: State clearly what the argument is—and is not.

- **AI's Role**: 1. Help condense a long analytical section into a precise and focused thesis statement. 2. Offer alternative formulations of the argument with variations in tone or scope, such as a bold versus a more cautious framing. 3. Suggest how the central claim intersects with broader historiographical debates, identifying relevant schools of thought or comparative cases.

*Prompt*: "Summarize my argument in one sentence. Then show how it contrasts with existing literature on Chinese elite philanthropy."

## 🟢 *Support Writing* – 2. Structure the Argument Logically

**Goal**: Break the claim into sub-arguments, each supported by evidence.

- **AI's Role**: 1. Help draft a scaffolding that maps claims to evidence clusters; 2. Identify logical gaps, missing transitions, or internal contradictions.

*Use case*: Ask AI to outline the progression from local case study → institutional dynamics → broader historiographical implications.

## 🔵 *Discover* – 3. Position the Argument Historiographically

**Goal**: Clarify how your interpretation engages with or departs from previous views.

- **AI's Role**: 1. Compare key passages to summary positions in historiography; 2. Suggest authors or schools whose views are relevant for dialogue or critique.

*Prompt*: "What historians should I cite if I want to challenge the idea that Shanghai merchants were primarily driven by self-interest?"

## 🟢 *Support Writing* – 4. Weave Together Source Types and Scales

**Goal**: Integrate qualitative and quantitative data, micro and macro analysis, or Chinese- and foreign-language sources.

- **AI's Role**: 1. Help transition between evidence types (e.g., from statistical tables to narrative analysis); 2. Suggest templates for integrating visuals (e.g., network diagrams, maps) into prose.

*Example*: "Write a paragraph that connects my topic model results to the narrative about merchant health policy."

###  *Support Writing* – 5. Anticipate Counterarguments

**Goal**: Strengthen the analysis by acknowledging potential objections or limitations.

- **AI's Role**: 1. Generate plausible counterarguments or alternate explanations from within or outside the field; 2. Suggest rhetorical strategies to address these without undermining the argument.

*Prompt*: "What are three plausible critiques of my argument that merchant philanthropy was politically strategic?"

###  *Visualize* – 6. Support Interpretive Claims with Visual Tools

**Goal**: Use diagrams, maps, or charts to anchor interpretive points.

- **AI's Role**: 1. Turn structured data (e.g., event timelines, name networks) into annotated visualizations; 2. Help identify patterns that could bolster narrative claims (e.g., clustering of events or actors).

*Example*: Network diagram showing overlapping membership in charitable associations and local councils.

## 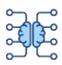 AI Affordances in Step 6

| Task | AI Function / Tool | Output for the Historian |
|------|-------------------|--------------------------|
| Formulate thesis | Argument distillation, phrasing | Clear, compelling core claim |
| Structure logic | Outline generation, logic checks | Argument broken into manageable, coherent segments |
| Historiographical positioning | Literature matching, school comparison | Argument placed in scholarly context |
| Source integration | Narrative linking, modality transition | Coherent mix of evidence types |
| Counterargument simulation | Alternate interpretations | Anticipatory framing and critical depth |
| Visual support | Diagram/text integration | Visual elements that support interpretation |

Once the interpretive skeleton is in place, it is time to bring the project to discursive life. Step 7 is the writing phase—where the historian drafts, refines, and polishes prose in a recursive and rhetorically sensitive process. AI can function here as an editorial assistant, supporting clarity, consistency, and narrative cohesion.

# Step 7: Write and Revise

 **Description**

This stage translates analytical work into a **narrative form** grounded in evidence, guided by historiography, and shaped by conceptual clarity. It is **iterative**, involving cycles of drafting, refining, and rethinking. Writing is not the endpoint of analysis but a space of historical interpretation, where framing, evidence, and voice are aligned. Key tasks include: 1. Formulating a compelling *problématique*. 2. Outlining sections anchored in central arguments. 3. Building a structured narrative through section drafting. 4. Ensuring coherence, clarity, and logical flow. 5. Cross-checking the use of evidence and data. 6. Integrating and enriching citations and references. 7. Crafting a strong introduction and conclusion.

**AI's Role:**

🟢 *Support Writing* **– 1. Formulate and Refine the *Problématique***

**Goal**: Articulate a central research question that unifies the work intellectually and thematically.

- **AI's Role**: 1. Provide multiple phrasings of the research problem based on a short project summary. 2. Compare similar formulations in existing literature, either through prompt-based comparison or the use of embedding tools. 3. Suggest counter-questions or alternative formulations to test the robustness and clarity of the proposed problématique.

*Prompt*: "Given this summary, what is a compelling historiographical problématique I could frame this article around?"

🟢 *Support Writing* **– 2. Draft a Structured Outline with Core Arguments**

**Goal**: Organize the work into sections aligned with analytical goals, not just chronology or themes.

- **AI's Role**: 1. Assist in creating an argument-centered outline (e.g., "List the main claims this article seems to make and map them into a 5-part structure."); 2. Test whether each section supports the central problématique.

*Outcome*: Structured outline with provisional section titles and embedded claims.

## 🟢 *Support Writing* – 3. Draft Section by Section (Not Linearly)

**Goal**: Build prose gradually, starting with the most developed section (often not the introduction).

- **AI's Role**: 1. Summarize complex notes into rough prose (e.g., turn bullet points on philanthropic networks into paragraph-level prose); 2. Offer stylistic refinement for clarity and scholarly tone.

*Use case*: Turn a paragraph of notes and quotes into an initial draft of a subsection.

## 🟢 *Support Writing* – 4. Check Argument Coherence and Logic

**Goal**: Ensure claims follow from evidence, and the narrative maintains internal logic and progression.

- **AI's Role**: 1. Analyze argument structure in a given section; 2. Highlight gaps in reasoning, missing transitions, or redundant passages; 3. Suggest rhetorical improvements for flow.

*Prompt*: "Review this section for internal consistency and argument development."

## 🔴 *Analyze* – 5. Verify Use of Evidence and Data

**Goal**: Confirm that each claim is grounded in primary or secondary sources and that data (quantitative or qualitative) is contextualized and cited.

- **AI's Role**: 1. Flag claims in the text that are not followed by supporting evidence. 2. Cross-check data tables, charts, or figures against the surrounding textual interpretation for consistency. 3. Provide reminders to contextualize statistics or visualizations within the broader narrative or argument.

*Outcome*: Evidence-accountability checklist.

### 🟢 *Support Writing* – 6. Enrich and Standardize References

**Goal**: Ensure thorough, accurate, and stylistically consistent citation of primary and secondary sources.

- **AI's Role**: 1. Suggest additional relevant references based on content and citations in similar works. 2. Format citations according to the required style, such as Chicago or APA. 3. Fill in missing metadata, including publication year, place, and publisher.

*Tool*: Zotero + GPT integration or manual prompt: "Generate full citation for this entry: Wang Xiaolai, 'On Modern Charity,' *Shenbao*, 1934

### 🟢 *Support Writing* – 7. Craft a Strong Introduction and Conclusion

**Goal**:

- The Introduction should present the problématique, justify the case, define scope, and preview the structure. The Conclusion should synthesize findings, return to the research question, and suggest implications or openings.
- **AI's Role**: 1. Review whether the introduction sets up the core questions and methods clearly. 2. Suggest stronger topic sentences or framing hooks to enhance engagement. 3. Help reframe the conclusion from a mere summary to a broader synthesis that highlights the significance of the findings.

*Prompt*: "Does this introduction effectively frame the significance of the argument? Suggest alternatives."

#  AI Affordances in Step 7

| Task | AI Method / Tool | Outcome for the Historian |
|---|---|---|
| Problématique refinement | LLM prompting, analogical generation | Sharpened research framing |
| Argument-based outlining | Summarization + structure prompts | Section plan anchored in analytical claims |
| Section drafting | Text expansion, paraphrasing | Prose generation from notes or bullets |
| Coherence and logic checking | Consistency checkers, logic prompts | Improved rhetorical flow and clarity |
| Data-evidence crosscheck | Gap detection, evidence mapping | Stronger linkage between argument and documentation |
| Reference enrichment | Citation completion + search tools | Complete and consistent bibliography |
| Introduction/conclusion refinement | Structural evaluation and rewriting | Engaging framing and meaningful closure |

The final stage begins once feedback is received—whether from peer reviewers, editors, or public readers. Step 8 involves a reflective return to the manuscript: rethinking, revising, and occasionally reframing the argument in light of critique. AI assists in organizing feedback, tracking changes, and clarifying language, helping the historian strengthen the work without losing their voice.

# Step 8: Respond, Revise, and Reframe (Post-Review Workflow)

 **Description**

After a manuscript is reviewed by peers or editors, the historian must: 1. Interpret critiques, which are often contradictory or uneven. 2. Assess which suggestions to accept, revise, or decline—and justify these decisions. 3. Rework structure, tone, references, or argument where necessary. 4. Compose a response letter that clearly maps changes to feedback. 5. Reflect on whether the revision clarifies, extends, or modifies the original research question or narrative. This process is also an opportunity to strengthen the manuscript's coherence, polish its framing, and sometimes open up new research avenues.

**AI's Role:**

🟠 *Analyze* **– 1. Digest and Organize Reviewer Comments**

**Goal**: Turn a messy block of comments into actionable tasks.

- **AI's Role**: 1. Summarize reviewer critiques into bullet points. 2. Cluster comments by theme, such as conceptual framing, clarity, or literature coverage. 3. Highlight contradictions between reviewers.

*Prompt*: "Here are the reviewer comments. Summarize them and group into categories for revision planning."

🔵 *Discover* **– 2. Strategize Revisions Based on Editorial Expectations**

**Goal**: Make principled decisions about what to change and what to defend, and how to explain it.

- **AI's Role**: 1. Suggest whether comments require structural revision, clarification, or simple insertion; 2. Simulate a potential editor's response to a planned revision or rebuttal.

*Prompt*: "This reviewer says my argument is too narrow. Should I broaden the scope or reframe the conclusion?"

 *Support Writing* – 3. Map Revisions onto the Manuscript

**Goal**: Systematically integrate changes throughout the text, ensuring consistency.

- **AI's Role**: 1. Highlight where specific comments should lead to changes (e.g., "Revise paragraph 2 in section 3 to address reviewer #2's critique on sources."); 2. Check for ripple effects of major conceptual changes.

*Outcome*: A dynamic to-do list or revision roadmap tied to page and section numbers.

 *Support Writing* – 4. Draft the Response Letter to Reviewers and Editors

**Goal**: Communicate clearly how each comment was addressed, including changes made, and rationale where suggestions were declined.

- **AI's Role**: 1. Generate formal but collegial response text for each point; 2. Cross-reference changes and suggest phrases for common challenges (e.g., "While we appreciate this suggestion, we have opted not to…").

*Prompt*: "Write a well-argued response explaining why I didn't include an additional case study, as requested by reviewer 1."

 *Support Writing* – 5. Revise for Framing, Coherence, and Tone

**Goal**: Ensure that changes enhance clarity and argument—not create inconsistency or patchwork prose.

- **AI's Role**: 1. Review updated sections for logical flow, tonal consistency, and stylistic alignment; 2. Suggest transitions that help integrate new content.

*Use case*: You have added a new paragraph on gender in a previously male-focused section—AI checks transitions and coherence.

 *Discover* – 6. Reassess the Problématique (if needed)

**Goal**: Determine whether revisions require updating the research question, scope, or claims.

- **AI's Role**: 1. Compare the revised introduction and conclusion to the original one; 2. Suggest whether new themes merit integration into framing or future work.

*Prompt*: "Has my revised manuscript drifted from the original problématique? Suggest how to refocus the introduction."

#  AI Affordances in Step 8

| Task | AI Function / Tool | Outcome for the Historian |
|---|---|---|
| Summarize and structure reviewer input | Comment clustering, summarization | Actionable revision checklist |
| Strategize revisions | Suggest scope/tone/content responses | Clarified revision priorities |
| Map changes to manuscript | Text-location tagging + revision tracking | Efficient and thorough implementation |
| Draft response letter | Polite, scholarly paraphrasing | Clear, professional communication with reviewers |
| Check revised coherence | Flow and logic checking | Integrated and stylistically unified text |
| Rethink framing | LLM comparison of intro/conclusion | Updated problem definition or clarification of aims |

With a complete manuscript, shift attention outward: how to share the work, in what formats, and with which audiences? Step 9 encompasses scholarly publication, public engagement, and the strategic dissemination of findings. AI tools facilitate this process by generating abstracts, social content, and visual accompaniments tailored for academic and general readers alike.

# Step 9: Disseminate and Engage

 **Description**

Historians may choose how to share, present, and publicize their findings across: 1. **Academic venues** such as journals, edited volumes, and university presses; 2. **Digital platforms** including repositories, institutional websites, and personal pages; 3. **Public-facing formats** like blogs, podcasts, media interviews, and social media; 4. **Interactive or data-driven formats** including visualizations, databases, maps, or digital exhibits.

The goals here are: 1. To ensure the work reaches relevant scholarly audiences; 2. To make research findable and interpretable via good metadata and summaries; 3. To engage wider publics through accessible language and open formats.

 **AI's Role:**

🟢 *Support Writing* **– 1. Generate Metadata, Abstracts, and Keywords**

**Goal**: Prepare submission-ready summaries and searchable metadata to aid discoverability and indexing.

- **AI's Role**: 1. Draft concise, structured abstracts based on article text. 2. Generate keywords by analyzing core concepts and terminology. 3. Propose titles optimized for clarity and searchability.

*Prompt*: "Write a 250-word abstract for this article, in academic tone, followed by 8 keywords."
*Use case*: Create metadata for repositories (e.g., HAL, Zenodo, institutional archives).

🟢 *Support Writing* **– 2. Format for Submission to Different Outlets**

**Goal**: Adapt the manuscript to specific editorial and formatting guidelines.

- **AI's Role**: 1. Check citation styles (Chicago, MLA, etc.) and adjust automatically. 2. Flag content that exceeds word limits or lacks required components (e.g., acknowledgements, figure captions). 3. Generate alternative article titles tailored to disciplinary vs. interdisciplinary journals.

*Prompt*: "Transform this Chicago-style bibliography into APA, and suggest a 15-word title for a political history journal."

## 🔴 *Translate & Contextualize* – 3. Produce Plain-Language Summaries and Media-Friendly Blurbs

**Goal**: Make the research accessible to non-specialists and support broader impact.

- **AI's Role**: 1. Translate dense academic language into plain English (or French, Chinese, etc.). 2. Generate summaries for websites, newsletters, or press releases. 3. Suggest headlines and excerpted quotes for public platforms.

*Prompt*: "Summarize this article in 100 words for a general audience interested in public health history."

## 🟢 *Support Writing* – 4. Draft Social Media Posts and Visual Content

**Goal**: Share findings with targeted networks (e.g., #twitterstorians, digital humanities forums, China scholars).

- **AI's Role**: 1. Generate tweet threads or LinkedIn posts summarizing main findings. 2. Suggest hashtags and tags based on academic subfields. 3. Create image captions or brief slides for conference teasers.

*Prompt*: "Draft a 5-tweet thread introducing my article on philanthropic networks in Republican-era Shanghai."

## 🟣 *Visualize* – 5. Support Interactive Outputs and Reuse

**Goal**: Publish accompanying materials such as datasets, visualizations, maps, or timelines.

- **AI's Role**: 1. Help transform structured data into graphs, networks, or annotated maps. 2. Generate tooltips, legends, or intro text for digital exhibits. 3. Suggest platforms for hosting, such as GitHub Pages, Scalar, Omeka, or Datawrapper.

*Example*: Create a dynamic timeline of hospital construction events extracted from a dataset of press clippings.

🟣 *Visualize* – **6. Publish Interactive Outputs and Repositories**

- **What it does**: Supports open access publication of visualizations, datasets, and exhibits.
- **How**: AI helps annotate visual content, generate descriptions, and optimize metadata.

*Example*: Upload a CSV of board memberships; AI helps generate a legend and summary for a public-facing Gephi graph.

#  AI Affordances in Step 9

| Task | AI Function / Tool | Output for the Historian |
|------|--------------------|--------------------------|
| Metadata generation | Abstract and keyword extraction | Submission-ready summaries and indexing tools |
| Format and submission adaptation | Style checking, formatting prompts | Outlet-specific versions of the manuscript |
| Public summaries | Plain-language conversion, tone adaptation | Texts for newsletters, blog posts, press releases |
| Social media dissemination | Thread generation, post planning | Posts for Twitter/X, LinkedIn, or Mastodon |
| Visualization and public data use | Chart or map scripting, labeling, platforms | Visual content and interactive components for readers |
| Engagement and response planning | Feedback summarization, counterarguments | Improved revisions and strategic responses |

# Case Study: "A Tale of Three Merchants"
# A Workflow in Practice

This visual workflow is based on my research paper, "A Tale of Three Merchants," which examines the intertwined professional, political, and philanthropic trajectories of three prominent Shanghai businessmen—Zhu Baosan, Yu Qiaqing, and Wang Xiaolai—between 1848 and 1949. It serves as a concrete example of how a historian can operationalize AI within a structured, flexible, and reflective research process. Whereas the first part of this paper presents an ideal-type model centered on LLM-supported tasks, the case study that follows moves beyond abstraction to foreground the practical complexities of real-life research. In this more grounded workflow, the historian—represented here as "Mind"—occupies the central position, orchestrating and critically evaluating outputs from both LLMs and traditional computational methods (e.g., NLP pipelines in R or Python). Rather than placing AI at the core, this model emphasizes the historian's agency in selecting, sequencing, and combining tools according to the epistemic needs of the project. It shows how LLMs can work in tandem with code-based approaches and human judgment, each contributing in distinct but complementary ways to a transparent, reproducible, and interpretively rich research process.[8] The workflow thus maps each phase of the research process (from literature review to data analysis and manuscript finalization) onto specific tasks assigned across three modalities: LLM (AI-assisted), Mind (historian-driven), and Computational (tools such as Python, R, or Cytoscape).

This triadic model clarifies the evolving role of AI in historical research—not as a replacement for scholarly labor, but as a force multiplier when embedded within an ecosystem of methodological rigor, reproducibility, and reflexivity.

## Overview of the Workflow

The workflow demonstrates an integration of three cognitive/computational approaches across 2 research phases and 7 research steps with 86 total tasks:

LLM Tasks: 30 (44.8%) - Primarily for extraction, analysis, and writing support

Mind Tasks: 30 (44.8%) - Critical thinking, curation, and domain expertise

---

[8] LLMs here refer to using ChatGPT (OpenAI) or Claude (Anthropic) through their prompt interface, including specific agents that I designer for various tasks.

: 18 (26.9%) - Data processing, modeling, and visualization. The contribution of each approach varies both quantitatively and qualitatively at each step (Figure 5).

**Figure 5. Respective Weight of Human Cognition, Computational Methods, and LLMs in the Case Study Workflow.**

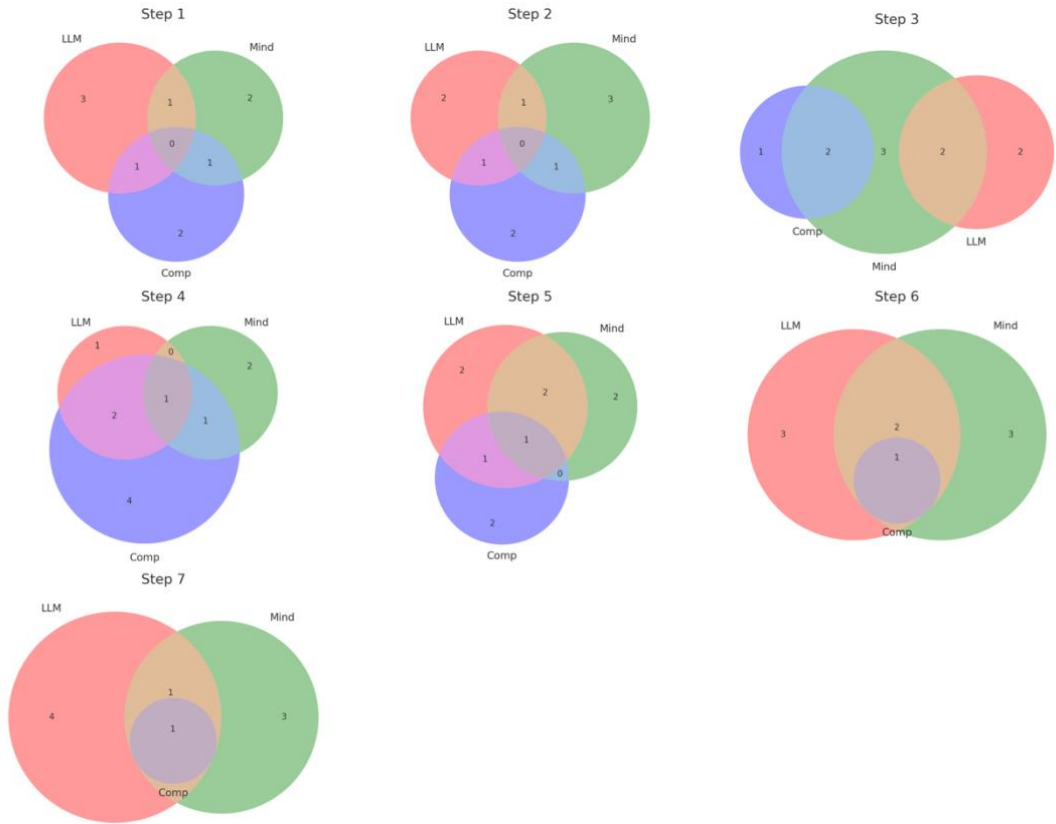

Note: This is a Venn diagram representation of the tabular data that describes the steps and tasks of the case study workflow (see AI Case Study Workflow on the GitHub repository). [Step 1: Explore a Historical Question; Step 2: Literature Review; Step 3: Define Scope and Methodology; Step 4: Locate/Collect Sources; Step 5: Analyze Sources & Data; Step 6: Buld an Argument; Step 7: Write and Revise] Produced with Claude Sonnet 4.

## Core Dimensions of a Hybrid Research Workflow

This project is characterized by four key dimensions that together define a robust and innovative research workflow (Figure 6).

**First**, methodological triangulation ensures both depth and reliability. The study relies on the effective combination of multiple approaches for validation and comparison. Topic modeling results are not only generated computationally but also interpreted through the complementary perspectives of both large language models (LLMs) and human expertise. Similarly, network analysis data is processed using computational tools but subjected to dual-layer interpretation by LLM and the historian alike. Throughout the process, multiple verification loops are embedded to guarantee accuracy and coherence across outputs.

**Second**, the integration of AI is purposeful and strategically targeted. LLMs are deployed in tasks where their strengths are most valuable: extracting names, organizations, and references from large corpora; assisting with the interpretation of topic modeling results and network data; and supporting writing-related activities such as copy editing, formatting verification, and content analysis. They are also employed as tools of quality control, particularly for checking references and verifying technical terminology.

**Third**, the workflow is grounded in computational rigor. A diverse set of specialized tools is brought to bear on different aspects of the research process: R is used for statistical analysis and text processing, Python supports data manipulation and visualization, and Cytoscape facilitates network analysis. Each step involves careful verification procedures, including cross-platform export and consistency checks, to ensure methodological soundness.

**Finally**, the workflow reflects a well-balanced pattern of human–AI collaboration. While AI systems are used to enhance scale and efficiency, human oversight remains central at all critical decision points. The historian intervenes at multiple levels: crafting precise prompts to guide LLM outputs, selecting relevant sources or segments for analysis, and reviewing AI-generated content—such as extracted entities, topic labels, or summaries—for accuracy, nuance, and contextual appropriateness. In tasks like topic modeling interpretation or network narrative construction, AI suggestions are treated as hypotheses to be verified, refined, or rejected, often paired with parallel implementations in R or Python. This collaborative model leverages the respective strengths of human judgment and machine processing, achieving results that are both scalable and interpretively grounded.

**Task Breakdown by Research Step**

In addition to the seven core research steps outlined in the AI & History workflow, this case study includes two framing phases—labeled Phase 0 ("Workflow Setup & Documentation") and Phase 00 ("Documentation & Reproducibility"). These do not correspond to stages of historical inquiry *per se* but instead address project management tasks that precede and follow the main research cycle. They reflect the conditions under which AI-supported historical research is initiated, sustained, and ultimately preserved.

**Figure 6.  The AI-Augmented Case Study Workflow**

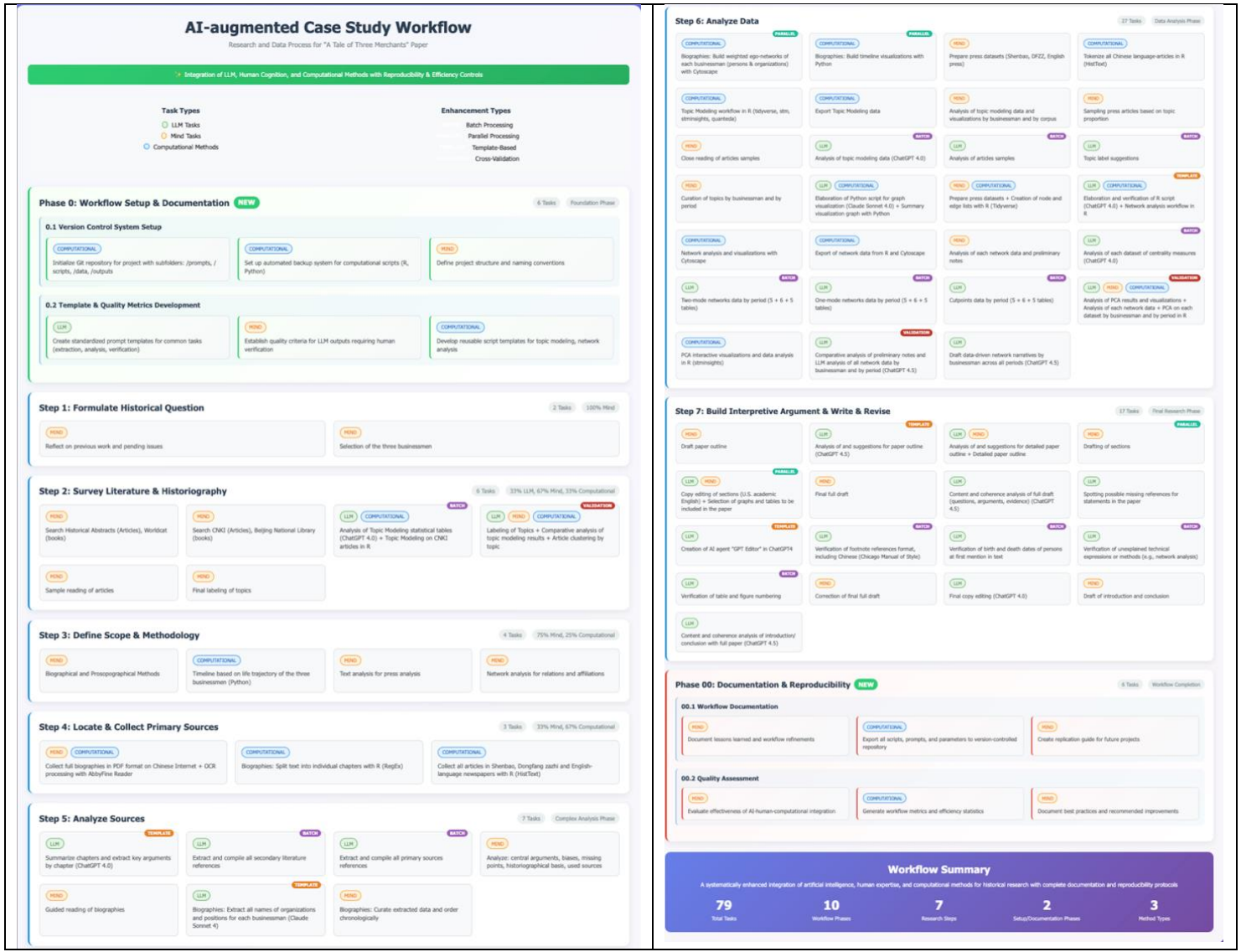

Note: Graph visualization produced with Claude Sonnet 4

## Reading the Workflow: Task Interactions and Interpretive Logic

In Figure 6, the overall logic remains the same: the research process unfolds sequentially and vertically, moving step by step. However, within each step, there is also a horizontal dynamic that connects tasks not only in sequence but also through iterations between modalities—for example, from computational methods to LLMs—and sometimes through tasks that run in parallel. In practical terms, a more faithful representation would resemble a *cross-modal iteration* between tasks.

For example, at Step 2 (Literature Review), after collecting the relevant papers, I first used my "GPT LitRev" agent to read and summarize the texts, extract key arguments, and identify primary and secondary sources [LLM]. I then turned to R to apply topic modeling using three different libraries: `stm` for analysis and basic visualization, `stminsights` for exploratory visualizations, and `LDAvis` for interactive visualization [Computational]. I conducted an initial analysis and topic labeling myself, but I also submitted the same statistical output to ChatGPT for topic analysis and labeling [Parallel Mind & LLM]. This process, grouped under "batch processing," enabled a comparative analysis of the LLM results and my own, which in turn helped validate the topics and cluster the articles by topic proportion [Mind & LLM & Computational]. For each topic, I read the articles with the highest proportional score to assess topic validity and fine-tune the topic labels [Mind]. This foundational work enabled me to write a detailed literature review, which then served as the basis for the more concise review included in the paper.

Step 6 exemplifies the most fully developed and complex articulation of Mind, Computational methods, and LLMs. Building on the extracted data from Step 5, I relied primarily on computational tools—over which I maintain full control (acknowledging that library-based algorithms also introduce their own constraints)—to produce the groundwork for timeline visualizations, construct ego-networks for each businessman, and apply topic modeling to the 30,000 press articles from *Shenbao*, *Dongfang zazhi*, and English-language newspapers. I began with my own analysis and article reading, and then repeated the topic modeling analysis using ChatGPT. This enabled comparative review and refinement of the topics.

For network analysis, I followed a similar pattern of alternating between computational methods and human reasoning. However, given the sheer volume of data, LLMs proved especially valuable in assisting with the interpretation of the statistical datasets generated by the network analyses. At the final stage, I introduced Principal Component Analysis (PCA) to gain a synthetic overview of the network data, using computational methods for calculation and visualization, and LLMs for backup interpretation. This integrated process laid the foundation for writing extensive, data-driven network narratives for each businessman, which then supported the more qualitative network analysis presented in the paper.

# Illustrative Examples of LLM Integration across Research Step

**Step 1: Formulate Historical Question**

While reflecting on Zhu Baosan's philanthropic activities, I used GPT-4 to rephrase an early research question into comparative formulations ("How did elite philanthropy function as a tool of civic governance in treaty-port Shanghai?"), then assessed these against historiographical coverage.

**Step 2: Survey Literature & Historiography**

After topic modeling over 300 CNKI articles related to merchants and philanthropy, I asked ChatGPT to label the resulting topics and compare them to human-labeled categories. Disagreements (e.g., over whether a topic reflected "public health" or "institutional welfare") guided closer manual inspection and final classification.

**Step 3: Define Scope & Methodology**

I wrote a conceptual outline of the project's biographical-network approach, and then prompted GPT-4 to identify potential methodological gaps. It flagged the uneven treatment of transnational affiliations, which I then addressed by incorporating English-language sources into the source base.

**Step 4: Locate & Collect**

I used R and HistText to scrape and clean all *Shenbao* articles mentioning Wang Xiaolai (1910–1949), while I asked Claude 4 to summarize Wang's biographical PDFs chapter by chapter, spotting inconsistencies in place names and affiliations that required manual correction.

**Step 5: Analyze Sources & Data**

I created topic models of *Shenbao* articles for each merchant. I asked GPT-4 to assign a label to each topic and then to generate a synthetic narrative describing thematic evolution over time. I compared this with my own periodized interpretations and used it as a "counterpoint" for triangulation.

**Step 6: Build Interpretive Argument**

After writing an outline of the argument around the "adaptive governance" of Shanghai by elite merchants, I prompted GPT-4 to question the logical flow of claims. Its feedback helped identify a gap in linking social network centrality metrics to actual institutional power—a gap I addressed through additional analysis.

**Step 7: Write & Revise**

After completing a full draft, I created a dedicated GPT-4 "Editor" agent, which ran consistency checks across citation formatting, topic sentences, and the labeling of figures. I compared its findings with my own checklist and used the overlap to finalize revisions.

# Detailed Workflow: A Case-Based Integration of AI, Human Cognition, and Computational Methods

**Phase 0: Workflow Setup & Documentation**

0.1 Version Control System Setup

Computational: Initialize Git repository for project with subfolders: /prompts, /scripts, /data, /outputs

Computational: Set up automated backup system for computational scripts (R, Python)

Mind: Define project structure and naming conventions

0.2 Template & Quality Metrics Development

LLM: Create standardized prompt templates for common tasks (extraction, analysis, verification)

Mind: Establish quality criteria for LLM outputs requiring human verification

Computational: Develop reusable script templates for topic modeling, network analysis

**Step 1: Formulate Historical Question**

1.1 Conceptual Development

Mind: Reflect on previous work and pending issues

Mind: Selection of the three businessmen

1.2 Documentation Setup

Mind: Document research questions and hypotheses in version-controlled format

Computational: Create project metadata file with research parameters

**Step 2: Survey Literature & Historiography**

2.1 Database Searches

Mind: Search Historical Abstracts (Articles), Worldcat (books)

Mind: Search CNKI (Articles), Beijing National Library (books)

2.2 BATCH Processing Implementation[9]

Mind: Design an LLM agent (GPT LitRev) in ChatGPT4 to pre-process documents for literature review

LLM: [BATCH] Analysis of Topic Modeling statistical tables (ChatGPT 4.0) - Group all statistical analysis tasks

Computational: Topic Modeling on CNKI articles in R with automated logging

Computational: Automated validation checks for topic modeling consistency

2.3 Cross-Validation Protocol

---

[9] [BATCH] refers to a mode of prompting large language models (LLMs) where multiple, similar tasks (e.g., extracting entities or labeling topics across many documents) are processed in a single, continuous session using standardized prompts. This approach improves consistency and efficiency across large datasets.

**LLM**: [BATCH] Labeling of Topics - Use standardized prompt template

**Mind**: Comparative analysis of topic modeling results

**Computational**: Article clustering by topic (by decreasing highest proportion)

**Computational**: Statistical comparison between **LLM** and computational topic assignments

2.4 Quality Control Loop

**Mind**: Sample reading of articles

**Mind**: Final labeling of topics

**Mind**: Document discrepancies between AI and human topic analysis

**Step 3: Define Scope & Methodology**

3.1 Methodological Framework

**Mind**: Biographical and Prosopographical Methods

**Computational**: Timeline based on life trajectory of the three businessmen (Python)

**Mind**: Text analysis for press analysis

**Mind**: Network analysis for relations and affiliations

3.2 API Integration Setup

**Computational**: Establish R/Python → **LLM** API connections for seamless data flow

**Computational**: Create automated data validation pipelines

**Step 4: Locate & Collect**

4.1 Data Collection with Validation

**Mind**: Collect full biographies in PDF format on Chinese Internet

**Computational**: Biographies: OCR processing with AbbyFine Reader

**Computational**: Automated OCR quality assessment and validation

4.2 Text Processing Pipeline

**Computational**: Biographies: Split text into individual chapters with R (RegEx)

**Computational**: Collect all articles in Shenbao, Dongfang zazhi and English-language newspapers with R (HistText)

4.3 Quality Assurance

**Computational**: Automated text segmentation validation

**Computational**: Cross-reference article counts across databases

**Step 5: Analyze Sources & Data**

5.0 Content Analysis with Validation

**LLM**: Summarize chapters and extract key arguments by chapter (ChatGPT 4.0) - Use standardized prompt template

**Computational**: Automated consistency checks for extracted summaries

**Mind**: Spot-check sample of **LLM** summaries against original text

5.1-5.2 Reference Extraction (**BATCH** Processing)

**LLM**: [BATCH] Extract and compile all secondary literature references - Standardized extraction template

LLM: [BATCH] Extract and compile all primary sources references - Same session as above

Computational: Automated deduplication and format validation of references

5.3-5.4 Source Analysis

Mind: Analyze: central arguments, biases, missing points, historiographical basis, used sources

Mind: Guided reading of biographies

5.5-5.6 Entity Extraction with Cross-Validation

Mind: Design an AI agent to extract biographical data (GPT Biodata) in ChatGPT4 and (Cld DataExtract) in Claude Sonnet 4.

LLM: Biographies: Extract all names of organizations and positions for each businessman (Claude Sonnet 4) - Template-based extraction

Mind: Biographies: Curate extracted data and order chronologically

Computational: Named Entity Recognition validation using computational NLP

Computational: Cross-validation between LLM and NLP extraction results

5.7-5.8 Network Construction with PARALLEL Processing[10]

Computational: [PARALLEL] Biographies: Build weighted ego-networks of each businessman (persons & organizations) with Cytoscape

Computational: [PARALLEL] Biographies: Build timeline visualizations with Python

Computational: Automated network topology validation

5.9-5.12 Press Analysis Pipeline

Mind: Prepare press datasets (Shenbao, DFZZ, English press)

Computational: Tokenize all Chinese language-articles in R (HistText)

Computational: Topic Modeling workflow in R (tidyverse, stm, stminsights, quanteda)

Computational: Export Topic Modeling data with automated metadata logging

5.13-5.17 Integrated Analysis with Cross-Validation

Mind: Analysis of topic modeling data and visualizations by businessman and by corpus

Mind: Sampling press articles based on topic proportion

Mind: Close reading of articles samples

LLM: [BATCH] Analysis of topic modeling data (ChatGPT 4.0) - Use BATCH processing for efficiency

LLM: [BATCH] Analysis of articles samples - Same session as above

LLM: [BATCH] Topic label suggestions - Same session as above

Mind: Curation of topics by businessman and by period

Computational: Statistical correlation analysis between human and LLM topic assessments

5.20-5.25 Network Analysis with API Integration

---

[10] [PARALLEL] indicates Computational or LLM tasks that were carried out simultaneously (but independently) on multiple datasets—e.g., generating networks for three individuals at once using the same script logic.

LLM: [API-INTEGRATED] Elaboration of Python script for graph visualization (Claude Sonnet 4.0)[11]

Computational: [API-INTEGRATED] Summary visualization graph with Python (pandas, matplotlib, numpy) - Direct data feed from R

Mind: Prepare press datasets (Shenbao, DFZZ, English press)

Computational: For each dataset and each businessman: Creation of node and edge lists with R (Tidyverse)

LLM: [TEMPLATE-BASED] Elaboration and verification of R script (ChatGPT 4.0)

Computational: 1. Network analysis workflow in R (lubridate, ggplot2, tidygraph, igraph)

Computational: 2. Network analysis and visualizations with Cytoscape

Computational: 3. Export of network data from R and Cytoscape with version control

5.26-5.34 Comparative Analysis with Systematic Cross-Validation

Mind: Analysis of each network data and preliminary notes

LLM: [BATCH] Analysis of each dataset of centrality measures (ChatGPT 4.0)

LLM: [BATCH] 1. Two-mode networks data by period (5 + 6 + 5 tables)

LLM: [BATCH] 2. One-mode networks data by period (5 + 6 + 5 tables)

LLM: [BATCH] 3. Cutpoints data by period (5 + 6 + 5 tables)

LLM: Analysis of PCA results and visualizations (ChatGPT 4.0)

Mind: Analysis of each network data and preliminary notes

Computational: PCA on each dataset by businessman and by period in R

Computational: PCA interactive visualizations and data analysis in R (stminsights)

Computational: Automated statistical validation of PCA results

LLM: [CROSS-VALIDATION] Comparative analysis of my preliminary notes and LLM analysis of all network data by businessman and by period (ChatGPT 4.5)

LLM: Draft data-driven network narratives by businessman across all periods (ChatGPT 4.5)

Mind: Systematic comparison and documentation of human vs. AI analytical insights

**Step 6: Build Interpretive Argument**

6.1-6.3 Argument Development with Version Control

Mind: Draft paper outline

LLM: Analysis of and suggestions for paper outline (ChatGPT 4.5) - Template-based feedback

LLM: Analysis of and suggestions for detailed paper outline

Mind: Detailed paper outline

Mind: Document version history of outline iterations

**Step 7: Write & Revise**

---

[11] [API-INTEGRATED] designates instances where LLMs were accessed through direct Application Programming Interface (API) calls from within a coding environment (e.g., R or Python), enabling dynamic, iterative interactions between AI and computational pipelines.

7.1-7.3 Drafting with PARALLEL Processing

Mind: [PARALLEL] Drafting of sections

LLM: [PARALLEL] Copy editing of sections (U.S. academic English) - BATCH process by section

Mind: [PARALLEL] Selection of graphs and tables to be included in the paper

Mind: Final full draft

7.4-7.6 Content Analysis and AI Agent Creation

LLM: Content and coherence analysis of full draft (questions, arguments, evidence) (ChatGPT 4.5)

LLM: Spotting possible missing references for statements in the paper

Mind: Creation of AI agent (GPT Editor) in ChatGPT4 for systematic verification

7.6-7.10 Systematic Verification (BATCH Processing)

LLM: [BATCH - GPT Editor] Verification of footnote references format, including Chinese (Chicago Manual of Style)

LLM: [BATCH - GPT Editor] Verification of birth and death dates of persons at first mention in text

LLM: [BATCH - GPT Editor] Verification of unexplained technical expressions or methods (e.g., network analysis)

LLM: [BATCH - GPT Editor] Verification of table and figure numbering

7.11-7.14 Final Integration with Quality Control

Mind: Correction of final full draft

LLM: Final copy editing (ChatGPT 4.0)

Mind: Draft of introduction and conclusion

LLM: Content and coherence analysis of introduction/conclusion with full paper (ChatGPT 4.5)

Computational: Automated consistency checking across all sections

**Phase 00: Documentation & Reproducibility**

00.1 Workflow Documentation

Mind: Document lessons learned and workflow refinements

Computational: Export all scripts, prompts, and parameters to version-controlled repository

Mind: Create replication guide for future projects

00.2 Quality Assessment

Mind: Evaluate effectiveness of AI-human-computational integration

Computational: Generate workflow metrics and efficiency statistics

Mind: Document best practices and recommended improvements

## Statement on AI-Assisted Development

This workflow was developed with sustained support from OpenAI's GPT-4 (ChatGPT), which served as a generative, analytical, and editorial partner throughout the process. The language model contributed at multiple levels: it assisted in conceptualizing a nine-step research protocol for historians working with AI, refining each step with typological tagging (Discover, Analyze, Support Writing, Visualize, Translate & Contextualize), and drafting both detailed descriptions and exercise-based workshop materials. It helped articulate ethical considerations around AI use, including citation practices, and generated visual summaries (e.g., matrix charts, Sankey diagrams, and case-based illustrations). The model also supported the editing, merging, and formatting of workflow documentation—including the synthesis of practical case-study materials derived from a research project on Shanghai merchants (1848–1949). It was further involved in building pedagogical tools such as student workbooks, slide decks, and interactive prompts. All AI-generated content was systematically reviewed, corrected, and expanded by the human author, who retained full interpretive responsibility for the structure, content, and scholarly framing of the final materials.

Claude Sonnet 4 (Anthropic) provided substantial assistance in the development and visualization of this enhanced AI-augmented historical research workflow. The AI's contributions included: (1) restructuring the original workflow from a linear 10-step process into a more logical framework with setup (Phase 0), core research steps (Steps 1-7), and documentation phases (Phase 00); (2) identifying and categorizing workflow enhancement opportunities through the systematic application of batch processing, parallel processing, template-based approaches, and cross-validation methodologies; (3) redistributing tasks between phases to create clearer functional divisions, particularly separating source analysis (Step 5) from data analysis (Step 6); (4) designing and implementing an interactive HTML visualization with color-coded phases, enhancement tags, and comprehensive task breakdowns; and (5) providing iterative refinements to ensure the workflow structure accurately reflected the integration of LLM, human cognition, and computational methods. The AI served as both a methodological consultant in workflow optimization and a technical collaborator in creating the visual documentation, demonstrating the recursive nature of AI-augmented academic work where the tool itself contributes to frameworks for its own scholarly application.